\documentclass[conference,10pt]{IEEEtran}

\usepackage{framed}
\usepackage{tcolorbox}
\usepackage{algorithm}
\usepackage[noend]{algpseudocode}
\usepackage{bm}
\usepackage{stackrel}
\usepackage{subfigure}
\usepackage{epsf}
\usepackage{amsmath,amssymb}
\usepackage{graphicx}
\usepackage{color}
\usepackage{cite}
\usepackage{multirow,tabularx}
\usepackage{ifthen}
\usepackage{epstopdf}
\usepackage{mathtools}
\input{epstopdf.sty}
\usepackage{times}

\input{epsf.sty}

\makeatletter
\def\BState{\State\hskip-\ALG@thistlm}
\makeatother

\newcommand{\hide}[1]{\ifthenelse{\boolean{false}}{#1}{}}

\newtheorem{theorem}{{\bf Theorem}}

\newtheorem{lemma}{{\bf Lemma}}

\newcommand{\qed}{\nobreak \ifvmode \relax \else
      \ifdim\lastskip<1.5em \hskip-\lastskip
      \hskip1.5em plus0em minus0.5em \fi \nobreak
      \vrule height0.75em width0.5em depth0.25em\fi}

\newcommand{\beq}{\begin{equation}}
\newcommand{\eeq}{\end{equation}}
\newcommand{\barr}{\begin{array}}
\newcommand{\earr}{\end{array}}

\newcommand{\benum}{\begin{enumerate}}
\newcommand{\eenum}{\end{enumerate}}

\newcommand{\bit}{\begin{itemize}}
\newcommand{\eit}{\end{itemize}}

\newcommand{\bc}{\begin{center}}
\newcommand{\ec}{\end{center}}

\newcommand{\bdes}{\begin{description}}
\newcommand{\edes}{\end{description}}

\newcommand{\bfig}{\begin{figure}}
\newcommand{\efig}{\end{figure}}

\newcommand{\bemq}{\begin{quote} \begin{em}}
\newcommand{\eemq}{\end{em} \end{quote}}

\newcommand{\bmp}{\begin{minipage}}
\newcommand{\emp}{\end{minipage}}

\newcommand{\bsp}{\begin{slide*}}
\newcommand{\esp}{\end{slide*}}
\newcommand{\bsl}{\begin{slide}}
\newcommand{\esl}{\end{slide}}

\newcommand{\blem}{\begin{lemma}}
\newcommand{\elem}{\end{lemma}}
\newcommand{\bthm}{\begin{theorem}}
\newcommand{\ethm}{\end{theorem}}

\newcommand{\EX}[1]{\mathbb{E}\left[ #1 \right]} 
\newcommand{\pr}[1]{\mathbb{P}\left[ #1 \right]}

\IEEEoverridecommandlockouts

\begin{document}

\title{Scheduling Policies for Age Minimization in Wireless Networks with Unknown Channel State} 
\author{Rajat Talak, Igor Kadota, Sertac Karaman, and Eytan Modiano
\thanks{The authors are with the Laboratory for Information and Decision Systems (LIDS) at the Massachusetts Institute of Technology (MIT), Cambridge, MA. {\tt \{talak, kadota, sertac, modiano\}@mit.edu} }
} 


\maketitle

\begin{abstract}
Age of information (AoI) is a recently proposed metric that measures the time elapsed since the generation of the last received information update.  We consider the problem of AoI minimization for a network under general interference constraints, and time varying channel. We study the case where the channel statistics are known, but the current channel state is unknown. We propose two scheduling policies, namely, the virtual queue based policy and age-based policy. In the virtual queue based policy, the scheduler schedules links with maximum weighted sum of the virtual queue lengths, while in the age-based policy, the scheduler schedules links with maximum weighted sum of a function of link AoI. We prove that the virtual queue based policy is peak age optimal, up to an additive constant, while the age-based policy is at most factor $4$ away from the optimal age. Numerical results suggest that both the proposed policies are, in fact, very close to the optimal.
\end{abstract}

\section{Introduction}
\label{sec:intro}
Age of information (AoI), at the destination node, is the time elapsed since the last received information update was generated at the source node. AoI, upon reception of a new update packet drops to the time elapsed since generation of the packet, and grows linearly otherwise. Unlike packet delay, AoI measures the lag in obtaining information at the destination node, and is therefore more suited for applications involving dissemination of time sensitive information.

AoI was recently proposed in~\cite{2011SeCON_Kaul, 2012Infocom_KaulYates}.
In~\cite{2011SeCON_Kaul}, AoI was studied for a network of vehicles exchanging status updates packets, via simulations, where it was shown that the AoI is minimized at a certain optimal packet generation rate. It was further shown that AoI can be improved by changing the queue discipline of the MAC layer FIFO queue to last-in-first-out (LIFO). This observation was proved under a general network setting in~\cite{BedewyISIT17_LIFO_opt}.
Motivated by~\cite{2011SeCON_Kaul}, AoI was analyzed for several queueing models~\cite{2012Infocom_KaulYates, 2015ISIT_LongBoEM, 2013ISIT_KamKomEp, 2016X_LongBo, 2012CISS_KaulYates, 2016ISIT_Najm, 2014ISIT_CostaEp}.

However, age minimization for a network under general interference constraints and time varying channels has received very little attention. A problem of scheduling finitely many update packets under physical interference constraints was shown to be NP-hard in~\cite{2016Ep_WiOpt}. Age for a broadcast network, where only a single link can be activated at any time, was studied in~\cite{2016allerton_IgorAge, 2017ISIT_YuPin}. Some preliminary analysis of age for a slotted ALOHA like random access was done in~\cite{2017X_KaulYates_AoI_ALOHA}, while age minimization under throughput constraints for a broadcast network, in which only a single link can be activated at a time, was only recently studied in~\cite{Igor18_infocom}.

In this paper, we considered the problem of age minimization for a wireless network under general interference constraints, and time varying channels. We consider active sources, which generate fresh information in every slot, and single-hop flows for which all source and destination nodes share a link. We propose two policies, namely, a virtual queue based policy $\pi_Q$ and an age-based policy $\pi_A$ that takes into account the AoI in making decisions. In the queue based policy $\pi_Q$, each link maintains a virtual queue, and a set of non-interfering links with the highest weighted sum of virtual queue lengths is scheduled, in every time slot. In the age-based policy, however, the set of non-interfering links with the highest weighted sum of link-AoIs are scheduled.

We show that the virtual-queue based policy is peak age optimal, up to an additive factor, and that the age-based policy is at most a factor $4$ away from the optimal peak and average age. A similar result was obtained for broadcast network, in which at most one link can be activated simultaneously, recently in~\cite{Igor18_infocom}.
Numerical simulations indicate that both the policies are very close to the optimal peak and average age, and outperform the stationary policy proposed in~\cite{talak17_StGenIC_Mobihoc}, especially when the network interference is high.

This is an extension of our recent work in~\cite{talak17_StGenIC_Mobihoc, talak17_WiOpt}, where we proposed similar policies when the channel states are perfectly known for each slot.


\section{System Model}
\label{sec:model}
The wireless network is modeled as a graph $G = (V, E)$, where $V$ denotes the set of nodes and $E$ the set of directed links. We consider a slotted time system, where the slot duration is normalized to unity. Due to wireless interference, not all links can be activated simultaneously. We call a set $m \subset E$ that can be activated simultaneously without interference as a \emph{feasible activation set}, and use $\mathcal{A}$ to denote the collection of all feasible activation sets.

We use $U_{e}(t)$, which equals either $0$ or $1$, to denote whether the link $e$ is activated or not, respectively, at time $t$.
Not every attempted transmission on a link is successful due to channel errors, and we use $S_{e}(t) \in \{0, 1\}$ to denote the state of link $e$ at time $t$. If $S_{e}(t) = 1$ then an attempted transmission over $e$ at time $t$ succeeds, and fails otherwise. A successful transmission occurs over link $e$, at time $t$, if and only if $U_{e}(t)S_{e}(t) = 1$.

We assume the channel process $\{ S_{e}(t) \}_{t\geq 0}$ to be independent and identically distributed (i.i.d.) across time, with $\gamma_e = \pr{S_{e}(t) = 1} > 0$ for all $e \in E$. Note that the channel is not identical across links $e$, and we allow for the channel success probability $\gamma_e$ to be different across links.

\begin{figure}
  \centering
  \includegraphics[width=0.95\linewidth]{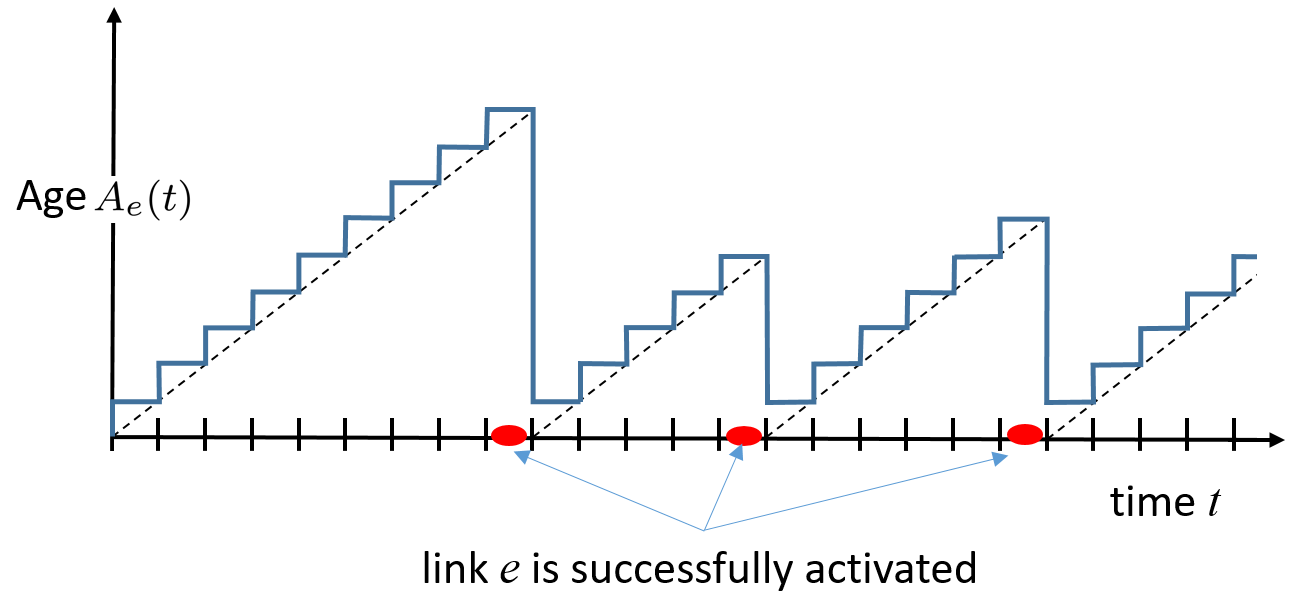}
  \caption{Evolution of age of link $e$, namely $A_{e}(t)$, as a function of time $t$.}\label{fig:age_evol}
\end{figure}

We consider \emph{active nodes}, which transmit fresh information at every transmission opportunity. We define age $A_{e}(t)$, of a link $e$ at time $t$, to be the time elapsed since the last successful activation of link $e$. Figure~\ref{fig:age_evol} shows evolution of age $A_{e}(t)$ for a link $e$. Age $A_{e}(t)$ reduces to $1$ upon a successful activation of link $e$, while it increases by $1$ in every slot in which there is no successful activation of link $e$. This age evolution equation can be written as
\begin{equation}\label{eq:age_evol2}
A_{e}(t+1) = 1 + A_{e}(t) - U_{e}(t)S_{e}(t)A_{e}(t),
\end{equation}
for all $e \in E$, and $t \geq 0$.

For a link $e$, we define average age to be the area under the age curve in Figure~\ref{fig:age_evol}. This can be written as
\begin{equation}
\overline{A}^{\text{ave}}_{e} = \limsup_{t \rightarrow \infty} \EX{\frac{1}{t}\sum_{\tau = 0}^{t-1}A_{e}(\tau)}.
\end{equation}
We see that the age curve in Figure~\ref{fig:age_evol} peaks whenever the link $e$ is successfully activated. We define peak age, for a link $e$, to be the average of all the peaks. This is given by
\begin{equation}
\label{eq:peak_def}
\overline{A}^{\text{p}}_{e} = \limsup_{t \rightarrow \infty} \frac{ \EX{\sum_{\tau = 0}^{t-1} U_{e}(\tau)S_{e}(\tau)A_{e}(\tau)} }{ \EX{\sum_{\tau = 0}^{t-1}U_{e}(\tau)S_{e}(\tau)} },
\end{equation}
because the numerator is the sum of age peaks until time $t$, while the denominator counts the number of age peaks until time $t$. Note that, this follows because the age peaks occur whenever link $e$ is successfully activated, which is exactly when $S_{e}(\tau)U_{e}(\tau) = 1$.

Average and peak age of the network $G$ is then defined as the weighted sum of link ages:
\begin{equation}
\overline{A}^{\text{ave}} = \sum_{e \in E}w_e \overline{A}^{\text{ave}}_{e}~~~\text{and}~~~\overline{A}^{\text{p}} = \sum_{e \in E}w_e \overline{A}^{\text{p}}_{e},
\end{equation}
where $w_e > 0$ are positive weights. We are interested in designing policies that minimize peak and average age of the network.


\subsection{Scheduling Policies}
\label{sec:scheduling_policies}
A scheduling policy determines the set of links $m_t \subset E$ that will be activated at each time $t$. We consider policies that can make use of current and past age, and previous decisions, when making the decision at time $t$, but not the current channel state. That is, the scheduler at each time $t$ determines $m_t$ as a function of the set
\begin{equation}
\mathcal{H}(t) = \{\mathbf{U}(\tau), \mathbf{A}(\tau')~|~0 \leq \tau < t,~0\leq \tau' \leq t \}.
\end{equation}
Note that knowledge of age until time $t$, and decisions until time $t-1$, implies complete knowledge of the channel state $\mathbf{S}(\tau)$ until time $\tau \leq t-1$. This is because age $A_{e}(\tau)$ drops whenever $S_{e}(\tau)U_{e}(\tau) = 1$, and continues to grow if $S_{e}(\tau) = 0$.
We consider centralized scheduling policies, in which this information $\mathcal{H}(t)$ is centrally available to a scheduler. This assumption is consistent with that in network scheduling literature~\cite{TassiEp92, Neely_book}.

For such a policy $\pi$ we define link activation frequency to be
\begin{equation}
f_{e}(\pi) = \lim_{t \rightarrow \infty} \frac{1}{t}\sum_{\tau = 0}^{t-1}\mathbb{I}_{\{e \in m_t, m_t \in \mathcal{A} \}},
\end{equation}
for all $e \in E$. Note that, $U_{e}(t) = \mathbb{I}_{ \{e \in m_t, m_t \in \mathcal{A} \} }$, and thus, the link activation frequency doesn't count the channel uncertainties. However, it is clear that if $f_{e}(\pi) = 0$ or not well defined, for some $e$, then the peak and average age would be infinity or not-well defined.
We consider $\Pi$ to be the class of all policies for which the link activation frequencies $\mathbf{f}(\pi)$ are well defined and positive:
\begin{equation}
\Pi = \left\{ \pi \big|~f_{e}(\pi)~\text{exists and is positive}~\right\}. \label{eq:Pi}
\end{equation}
We define optimal peak and average age to be
\begin{equation}
\overline{A}^{\text{p}\ast} = \min_{\pi \in \Pi} \overline{A}^{\text{p}}(\pi)~~~~\text{and}~~~~\overline{A}^{\text{ave}\ast} = \min_{\pi \in \Pi} \overline{A}^{\text{ave}}(\pi).
\end{equation}

An important fact about the policy space $\Pi$ is that the space of all feasible link activation frequencies $\mathcal{F} = \left\{ \mathbf{f}(\pi)~|~\pi \in \Pi\right\}$, is given by
\begin{equation}
\mathcal{F} = \left\{ \mathbf{f} \in \mathbb{R}^{|E|} \Big| \mathbf{f} = M \mathbf{x},~\mathbf{1}^{T}\mathbf{x} \leq 1,~\text{and}~\mathbf{x} \geq 0 \right\}, \label{eq:mathcalF}
\end{equation}
where $M$ is a $|E|\times |\mathcal{A}|$ matrix such that $M_{e,m} = 1$ if and only if $e \in m$, and $0$ otherwise, for all $e \in E$ and $m \in \mathcal{A}$; see~\cite{Neely_book, TassiEp92}.

\subsection{Stationary Policies}
\label{sec:review}
An important sub-space of $\Pi$, which do not use any past history, is the space of stationary policies. In it, a feasible activation set $m \in \mathcal{A}$ is activated with probability $x_m$, in every slot; we have $\sum_{m \in \mathcal{A}} x_m = 1$. The link activation frequencies, for this policy, are then given by $f_e = \sum_{m: e \in m} x_m$, which can be written as $\mathbf{f} = M\mathbf{x}$, where $M$ is the same $|E|\times |\mathcal{A}|$ matrix in~\eqref{eq:mathcalF}. Therefore, any link activation frequency in the set $\mathcal{F}$, in~\eqref{eq:mathcalF}, can be achieved by a stationary policy.

The following result proves that there exists a stationary policy that is peak age optimal. This was proved in~\cite{talak17_StGenIC_Mobihoc}, and we will use it to prove bounds on our proposed policies.
\begin{framed}
\begin{theorem}
\label{thm:stationary}
The optimal peak age $\overline{A}^{\text{p}\ast}$ is given by
\begin{align}\label{eq:peak_age_opt}
\begin{aligned}
\overline{A}^{\text{p}\ast} =& \underset{\mathbf{f}, \mathbf{x} \in [0,1]^{|\mathcal{A}|}}{\text{Minimize}}
& & \sum_{e \in E} \frac{w_{e}}{\gamma_e f_{e}}, \\
& \text{subject to} && \mathbf{f} = M\mathbf{x}, \\
&&& \sum_{m \in \mathcal{A}} x_m \leq 1,
\end{aligned}
\end{align}
and the solution $\mathbf{x}^{\ast}$ to~\eqref{eq:peak_age_opt} yields a stationary policy that is peak age optimal, call it $\pi_C$. Furthermore, the peak age and average age for $\pi_C$ are equal and bounded by
\begin{equation}\label{eq:lo}
\overline{A}^{\text{p}\ast} = \overline{A}^{\text{p}}(\pi_C) = \overline{A}^{\text{ave}}(\pi_C) \leq 2\overline{A}^{\text{ave}\ast} - \sum_{e \in E}w_e.
\end{equation}
\end{theorem}
\end{framed}
\begin{IEEEproof}
This result is proved in our recent work~\cite{talak17_StGenIC_Mobihoc}.
To intuitively see the result, note that for a stationary policy with distribution $\mathbf{x}$ and link activation frequencies $\mathbf{f} = M\mathbf{x}$, every link $e$ is successfully activated  with probability $\gamma_e f_e$ in every slot. As a result, the age $A_{e}(t)$, which is the time since last activation is geometrically distributed with mean $\frac{1}{\gamma_e f_e}$. It turns out that the peak age of the link, is indeed, given by $\overline{A}^{\text{p}}_{e} = \frac{1}{\gamma_e f_e}$. As a consequence, the peak age of the network is given by $\overline{A}^{\text{p}} = \sum_{e \in E} \frac{w_e}{\gamma_e f_e}$, and the optimal peak age is given by~\eqref{eq:peak_age_opt}.

Furthermore,~\eqref{eq:lo} primarily follows because the peak and average age are equal for any stationary policy~\cite{talak17_StGenIC_Mobihoc}.
\end{IEEEproof}

%
A consequence of Theorem~\ref{thm:stationary} is that the stationary peak age optimal policy $\pi_C$ is also factor $2$ average age optimal. This can be seen from~\eqref{eq:lo}. Further, this bound is tight in the sense that for certain networks, average age of the stationary policy $\pi_C$ is indeed factor $2$ away from optimality. 

To see this, consider $E$ links only one of which can be activated at any given time $t$. Also, assume there to be no channel uncertainties, i.e. $S_{e}(t) = 1$ for all $t$ and $e$. If weights are all equal, i.e. $w_e = 1$ for all $e$, then the optimal stationary policy $\pi_C$ activates link $e$ with probability $1/|E|$. The average and peak age then is given by
\begin{equation}
A^{\text{ave}}(\pi_C) = A^{\text{p}}(\pi_C) = |E|^2.
\end{equation}
However, if we schedule links in $E$ in round robbin manner, the peak age would still be $|E|^2$ but the average age would improve to $|E|\left( |E| - 1 \right)/2$.

The above example shows that beyond stationary policies, average age could be improved by resorting to periodic policies.
One way to resort to periodic policies is to schedule links based on age $A_{e}(t)$. For example, the above round robbin policy can be induced by having the link with the largest age $A_{e}(t)$, or the largest $g\left( A_{e}(t)\right)$ for an increasing function $g(\cdot)$, transmit in every slot $t$. In this paper, we propose policies which do just that.
%

In Sections~\ref{sec:virtualQ-policy}, we propose a virtual queue based policy, which schedules a feasible activation set $m$ with largest weighted virtual queue lengths, rather than largest age.
In Section~\ref{sec:age-square-policy}, we propose an age-based policy that schedules feasible activation sets $m$ with maximum $\sum_{e \in m} w_e \gamma_e g\left( A_{e}(t)\right)$, for $g(x) = x^2 + \beta x$.

\section{Virtual-Queue Based Policy}
\label{sec:virtualQ-policy}
We first present a lemma that states a conservation law for age. Intuitively, it states that for any policy $\pi \in \Pi$, the sum of all age peaks is equal to the total time elapsed plus a small insignificant term that goes to $0$ as $t \rightarrow \infty$.
\begin{framed}
\begin{lemma}
\label{lem:time_conservation}
For any policy $\pi \in \Pi$ we have
\begin{equation}
\lim_{t \rightarrow \infty} \EX{\frac{1}{t}\sum_{\tau = 0}^{t-1}U_{e}(\tau)S_{e}(t)A_{e}(\tau)} = 1,
\end{equation}
for all $e \in E$.
\end{lemma}
\end{framed}
\begin{IEEEproof}
See Appendix~\ref{pf:lem:time_conservation}.
\end{IEEEproof}

A direct consequence of Lemma~\ref{lem:time_conservation} is that the peak age minimization problem $\min_{\pi \in \Pi} \overline{A}^{\text{p}}(\pi)$ reduces to
\begin{align}
\label{eq:peak_min_problem}
\begin{aligned}
& \underset{\bm{\alpha} \geq 0, \pi \in \Pi}{\text{Minimize}} & & \sum_{e \in E}\frac{w_e}{\alpha_e}, \\
& \text{subject to} && \liminf_{t \rightarrow \infty} \EX{\frac{1}{t}\sum_{\tau = 0}^{t-1}U_{e}(\tau)} \geq \frac{\alpha_e}{\gamma_e}~~\forall~e \in E.
\end{aligned}
\end{align}
We prove this equivalence in Appendix~\ref{pf:der_peak_min_prob}. This result is significant because it shows that the peak age minimization problem is independent of the age evolution equation. This is the reason why peak age minimization problem is much simpler than minimizing average age.

We now propose a policy that solves the peak age minimization problem~\eqref{eq:peak_min_problem}. Note that a policy $\pi$ can decide on the activation set $m_t$ at time $t$ based on the entire history $\mathcal{H}(t)$. However, we do not need the entire history to make a choice at time $t$ but only a representation of it.

To do so, we construct a virtual queue $Q_{e}(t)$, which is reduced by at most $1$ upon a successful transmission over link $e$ and increased otherwise. These queue lengths determine the `value' of scheduling link $e$ in time slot $t$. Therefore, a set $m_t \in \mathcal{A}$ that maximizes $\sum_{e \in m} w_e \gamma_e Q_{e}(t)$ is activated in slot $t$.
This virtual-queue based policy, $\pi_{Q}$, is described below. Here, $V > 0$ is any chosen constant.
\begin{framed}
\textbf{Virtual Queue based policy $\pi_Q$}
Start with $Q_{e}(0) = 1$ for all $e \in E$. At time $t$,
\begin{enumerate}
  \item Update $Q_{e}(t)$ as
        \begin{multline}
        Q_{e}(t) = \Bigg[Q_{e}(t-1) + \sqrt{\frac{V}{Q_{e}(t-1)}} \\ - S_{e}(t-1) U_{e}(t-1)\Bigg]_{+1}, \nonumber
        \end{multline}
        for all $e \in E$, where $[x]_{+1} = \max\{x, 1\}$.
  \item Schedule activation set $m_t$ given by
        \begin{equation}
        m_{t} = \arg\max_{m \in \mathcal{A}} \sum_{e \in m} w_e \gamma_e Q_{e}(t).
        \end{equation}
\end{enumerate}
\end{framed}
Note that, we require the product $U_{e}(t-1)S_{e}(t-1)$ at time $t$ in order to update the virtual queue lengths. This is possible with $\mathcal{H}(t)$, as the products $U_{e}(t-1)S_{e}(t-1)$ can be inferred from current age vector and past actions.
We now prove that the policy $\pi_Q$ is nearly peak age optimal up to an additive factor.
\begin{framed}
\begin{theorem}
\label{thm:piQ}
The peak age under for policy $\pi_Q$ is bounded by
\begin{equation}
\label{eq:thm:piQ}
\overline{A}^{\text{p}}(\pi_{Q}) \leq \overline{A}^{\text{p}\ast} + \frac{1}{2}\sum_{e \in E} w_e + \frac{1}{2V}\sum_{e \in E} w_e,
\end{equation}
where $\overline{A}^{\text{p}\ast}$ is the optimal value of~\eqref{eq:peak_age_opt}.
\end{theorem}
\end{framed}
\begin{IEEEproof}
Let $\alpha_{e}(t) = \sqrt{\frac{V}{Q_{e}(t)}}$ and $\overline{\alpha}_{e}(t) = \frac{1}{t}\sum_{\tau = 0}^{t-1}\alpha_{e}(\tau)$ for all $t \geq 0$ and $e \in E$. Also, let $g(\mathbf{\alpha}) = \sum_{e \in E} \frac{w_e}{\alpha_e}$ be the objective function in our optimization problem~\eqref{eq:peak_min_problem}.

The proof is divided into three parts, the proofs of which are given in Appendix~\ref{pf:thm:piQ}.

\textbf{Part A}: For all time $t$, we have
        \begin{equation}\label{eq:DPP_piQ_opt}
        \limsup_{t \rightarrow \infty} \EX{g\left( \overline{\bm{\alpha}}(t)\right)} \leq \overline{A}^{\text{p}\ast} + \frac{1}{2}\sum_{e \in E}w_e + \frac{1}{2V}\sum_{e \in E}w_e.
        \end{equation}

\textbf{Part B}: The virtual queue $\mathbf{Q}(t)$ is mean rate stable, i.e., for all $e \in E$ we have
        \begin{equation}
        \label{eq:b1}
        \limsup_{t \rightarrow \infty} \frac{1}{t}\EX{Q_{e}(t)} = 0.
        \end{equation}

\textbf{Part C}: If $\mathbf{Q}(t)$ is mean rate stable then
        \begin{equation}
        \label{eq:c1}
        \frac{1}{\gamma_e}\liminf_{t \rightarrow \infty} \EX{\overline{\alpha}_{e}(t)} \leq \liminf_{t \rightarrow \infty} \frac{1}{t}\EX{\sum_{\tau=0}^{t-1}U_{e}(\tau)},
        \end{equation}
        and
        \begin{equation}
        \label{eq:c2}
        \overline{A}^{\text{p}}\left( \pi_Q\right) \leq \limsup_{t \rightarrow \infty} \EX{g\left( \overline{\bm{\alpha}}(t)\right)}.
        \end{equation}

Since the virtual queues are mean rate stable, by Part~B,~\eqref{eq:c1} and~\eqref{eq:c2} are true. From~\eqref{eq:DPP_piQ_opt} and~\eqref{eq:c2} we get the result in~\eqref{eq:thm:piQ}.

Further, if we set
\begin{equation}
\alpha^{V}_{e} = \liminf_{t \rightarrow \infty} \EX{\overline{\alpha}_{e}(t)},
\end{equation}
for each $e \in E$, then $\bm{\alpha}^{V}$, with policy $\pi_Q$, solves the optimization problem~\eqref{eq:peak_min_problem}, up to an additive factor. To see this, notice that from~\eqref{eq:c1}, we know that $\bm{\alpha}^{V}$ satisfies the inequality constraint in~\eqref{eq:peak_min_problem}. Now, consider the objective function evaluated at $\bm{\alpha}^{V}$:
\begin{align}
g\left( \bm{\alpha}^{V} \right) &= g\left( \liminf_{t \rightarrow \infty} \EX{\overline{\bm{\alpha}}(t)} \right), \nonumber \\
&= \limsup_{t \rightarrow \infty} g\left(\EX{\overline{\bm{\alpha}}(t)} \right), \nonumber \\
&\leq \limsup_{t \rightarrow \infty} \EX{ g\left( \overline{\bm{\alpha}}(t) \right) }, \label{eq:oyo}
\end{align}
where the first equality is because $g$ is a continuous decreasing function in $\bm{\alpha}$, while the second inequality follows directly from Jensen's inequality as $g$ is convex.
Substituting~\eqref{eq:DPP_piQ_opt} in~\eqref{eq:oyo} we get
\begin{equation}
g\left( \bm{\alpha}^{V} \right) \leq \overline{A}^{\text{p}\ast} + \frac{1}{2}\sum_{e \in E}w_e + \frac{1}{2V}\sum_{e \in E}w_e.
\end{equation}
\end{IEEEproof}

Theorem~\ref{thm:piQ} shows that even when the channel statistics are not known the optimal peak age $A^{\text{p}\ast} = \sum_{e \in E}\frac{w_e}{\gamma_e f^{\ast}_{e}}$ can be achieved, barring an additive factor of $\frac{1}{2}\sum_{e \in E}w_e$, with arbitrary precision. The precision can be chosen by selecting $V$. For example, we may obtain peak age of at most $\overline{A}^{\text{p}\ast} + \frac{1}{2}\sum_{e \in E}w_e + \epsilon$ by setting $V = \frac{1}{2\epsilon}\sum_{e \in E}w_e$.
%



\section{Age-Based Policy}
\label{sec:age-square-policy}
In this section, we propose an age-based policy to minimize age of the network. To gain an intuitive understanding of the proposed policy, we first provide for an equivalent characterization of average age. Note that the average age for a link $e$ is given by $\overline{A}^{\text{ave}}_{e} = \limsup_{t \rightarrow \infty} \EX{\frac{1}{t}\sum_{t=0}^{t-1}A_{e}(\tau)}.$
The following result provides a different characterization of the average age in terms of $A_{e}^{2}(t)$.
\begin{framed}
\begin{lemma}
\label{lem:ave_age_conservation}
For any $\pi \in \Pi$, we have
\begin{multline}
\overline{A}^{\text{ave}}_{e} = \frac{1}{2}\limsup_{t \rightarrow \infty} \EX{\frac{1}{t}\sum_{\tau = 0}^{t-1} \gamma_e U_{e}(\tau) B_{e}(\tau)} + \frac{1-\beta}{2}, \nonumber
\end{multline}
for all $e \in E$, where $B_{e}(t) = A^{2}_{e}(t) + \beta A_{e}(t)$ and $\beta \in \mathbb{R}$.
\end{lemma}
\end{framed}
\begin{IEEEproof}
See Appendix~\ref{pf:lem:ave_age_conservation}.
\end{IEEEproof}
For an intuitive understanding of Lemma~\ref{lem:ave_age_conservation}, note that average age is essentially time averaged area of the triangles formed by the age curve in Figure~\ref{fig:age_evol}. Note that $S_{e}(t)U_{e}(t)A^{2}_{e}(t)$ are square of age peaks in the age curve Figure~\ref{fig:age_evol}, and $\frac{1}{2}S_{e}(t)U_{e}(t)A^{2}_{e}(t)$ is therefore the area of the triangle, because $S_{e}(t)U_{e}(t) = 1$ only at the instances when there is a successful transmission on link $e$. We can replace $S_{e}(t)$ with $\gamma_e$ because it is independent of $U_{e}(t)$ and $A_{e}(t)$. An additional term of $\beta A_{e}(t)$ is possible due to Lemma~\ref{lem:time_conservation}.

Lemma~\ref{lem:ave_age_conservation} implies that average age minimization problem over $\pi \in \Pi$ can be equivalently posed to minimize
\begin{equation}\label{eq:ave_age_proxy}
\limsup_{t \rightarrow \infty} \EX{\frac{1}{t}\sum_{\tau = 0}^{t-1}\sum_{e \in E} w_e \gamma_e U_{e}(\tau)B_{e}(\tau)}.
\end{equation}
Since, age reduces to $1$ after a link activation, it makes intuitive sense to choose $\mathbf{U}(t)$ such that as
\begin{equation}
\label{eq:ave_age_decision}
\mathbf{U}(t) = \arg\max_{\mathbf{U}^{'}(t)} \sum_{e \in E} w_e \gamma_e U^{'}_{e}(t)\left[ A^{2}_{e}(t) + \beta A_{e}(t)\right],
\end{equation}
in time slot $t$. This, in the least, should minimize age in the next slot. We now propose this age-based policy:
\begin{framed}
\textbf{Age-based Policy $\pi_{A}$}
The policy activates links $m_{t} \in \mathcal{A}$ in slot $t$ given by: \begin{equation}
m_{t} = \arg\max_{m \in \mathcal{A}} \sum_{e \in m} w_e \gamma_e \left[A^{2}_{e}(t) + \beta A_{e}(t)\right],
\end{equation}
for all $t \geq 1$.
\end{framed}
The following result shows that the average and peak age of policy $\pi_A$ is within a factor of $4$ from the respective optimal. 
\begin{framed}
\begin{theorem}
\label{thm:piA}
The policy $\pi_A$ is at most factor-4 peak and average age optimal, i.e.,
\begin{equation}
\label{eq:thm:piA_1}
\overline{A}^{\text{ave}}(\pi_A) \leq 4\overline{A}^{\text{ave}\ast} - c_1(\beta)\sum_{e \in E} w_e,
\end{equation}
and
\begin{equation}
\label{eq:thm:piA_2}
\overline{A}^{\text{p}}(\pi_A) \leq 4\overline{A}^{\text{p}\ast} - c_2(\beta)\sum_{e \in E}w_e,
\end{equation}
where $c_1(\beta) = \frac{10 + 2\beta - \beta^2}{4}$ and $c_2(\beta) = \frac{4 + 2\beta - \beta^2}{2}$.
\end{theorem}
\end{framed}
\begin{IEEEproof}
The proof is given in Appendix~\ref{pf:thm:piA}.
\end{IEEEproof}
We note that $\beta$ can be chosen to improve the additive factor of optimality. The best bounds, for both peak and average age, occur when $\beta = 1$, for which both $c_1(\beta)$ and $c_{2}(\beta)$ are maximized.

\section{Numerical Results}
\label{sec:numerical}
We now evaluate the performance of the proposed policies. We consider a $N = 20$ link network, with interference constraints such that at most $K$ links can be activated at any given time. We set link weights to unity, i.e., $w_e = 1$ for all $e \in E$. We let the links to be either `good', with channel success probability $\gamma_e = \gamma_{\text{good}} = 0.9$, or `bad' with channel success probability $\gamma_e = \gamma_{\text{bad}} = 0.1$. We use $\theta$ to denote the fraction of bad links in the network. We simulate the policies $\pi_Q$, $\pi_A$, and the peak age optimal stationary policy $\pi_C$ of~\cite{talak17_StGenIC_Mobihoc}, over a horizon of $10^5$ time slots.

\begin{figure}
  \centering
  \includegraphics[width=0.85\linewidth]{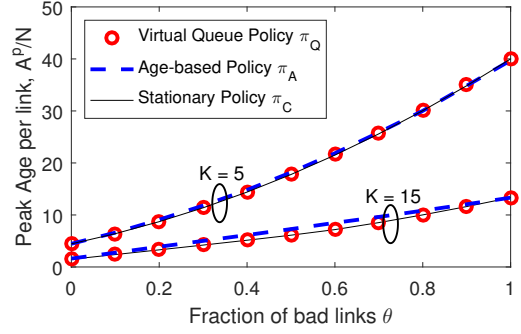}
  \caption{Peak age per link, namely $A^{\text{p}}/N$, as a function of the fraction of bad nodes $\theta$, for various policies.}\label{fig:main12peak}
\end{figure}

\begin{figure}
  \centering
  \includegraphics[width=0.85\linewidth]{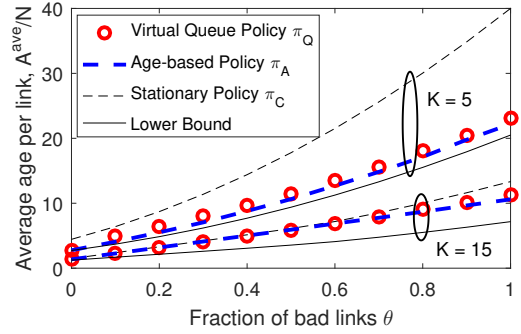}
  \caption{Average age per link, namely $A^{\text{ave}}/N$, as a function of the fraction of bad nodes $\theta$, for various policies.}\label{fig:main12ave}
\end{figure}

We first set $V = 1$ for policy $\pi_Q$ and $\beta = 1$ for policy $\pi_A$, and evaluate the policies. In figures~\ref{fig:main12peak} and~\ref{fig:main12ave} we plot the per-link peak and average age, $A^{\text{p}}(\pi)/N$ and $A^{\text{ave}}(\pi)/N$, respectively, for all policies $\pi \in \{\pi_Q, \pi_A, \pi_C \}$. As to be expected, we see that, increasing the fraction of `bad' channels $\theta$ or increasing interference, i.e. reducing $K$, increases age.

In Figure~\ref{fig:main12peak}, we further see that in all the cases, of $\theta$ and $K$, the peak age of the proposed policies $\pi_Q$ and $\pi_A$ coincide with the peak age optimal stationary policy $\pi_C$. Thus, the proposed policies are nearly, if not, peak age optimal under the current network setting. We observe similar behavior for several other networks, not presented here.


In Figure~\ref{fig:main12ave}, however, we observe a larger gap between average age of policy $\pi_C$ and proposed policies $\pi_Q$ and $\pi_A$, especially when $K = 5$ than when $K = 15$. This shows that the proposed policies perform much better than the stationary peak age optimal policy $\pi_C$ under high interference (smaller $K$). In Figure~\ref{fig:main12ave}, we also plot the average age lower bound, obtained from~\eqref{eq:lo}. We observe that the proposed schemes are much closer to the average age lower bound in the high interference case (small $K$), than in the low interference case (high $K$). This also shows that the age-based policy $\pi_A$ performs better than the bound derived in Theorem~\ref{thm:piA}. We believe that better bounds on $A^{\text{p}}(\pi_A)$ and $A^{\text{ave}}(\pi_A)$ are possible.

\subsection{Choice of Parameters $V$ and $\beta$}
\begin{figure}
  \centering
  \includegraphics[width=0.85\linewidth]{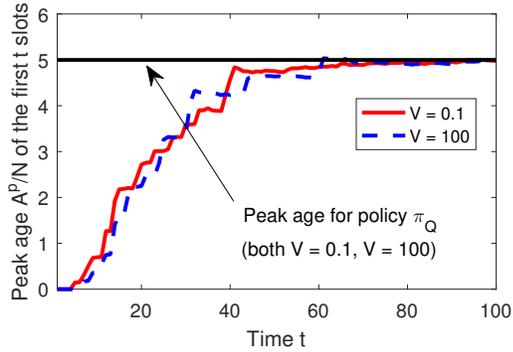}
  \caption{Peak age per link $A^{\text{p}}(\pi_Q)/N$, computed for the first $t$ slots, as a function of time $t$, for $V = 0.1$ and $V = 100$. Also plotted is the per-link peak age $A^{\text{p}}(\pi_Q)/N$ achieved over a much larger time horizon.}\label{fig:main14}
\end{figure}

\begin{figure}
  \centering
  \includegraphics[width=0.85\linewidth]{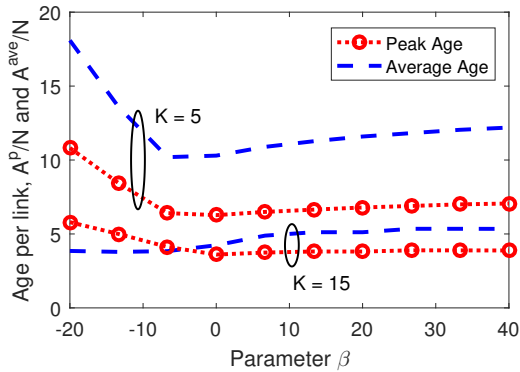}
  \caption{Per link peak and average age for policy $\pi_A$, namely $A^{\text{p}}(\pi_A)/N$ and $A^{\text{ave}}(\pi_A)/N$, as a function of parameter $\beta$.}\label{fig:main13}
\end{figure}

The virtual queue based policy $\pi_Q$ and the age-based policy $\pi_A$ have free parameters $V$ and $\beta$, respectively, which need to be chosen apriori. 
Figure~\ref{fig:main14} plots the per-link peak age $A^{\text{p}}/N$, computed over the first $t$ slots, as a function of time $t$. We observe that the choice of $V$ has nearly no effect on the convergence time of the algorithm. We observe that for $V$ as small as $0.1$ and as large as $V = 100$, the convergence time of peak age is similar.
In Figure~\ref{fig:main13}, we plot the per-link peak and average age for the age-based policy $\pi_A$, when $K = 5, 15$ and $\theta = 0.25$, i.e. when $25\%$ of the links are `bad'. We observe that the achieved age degrades dramatically for $\beta < 0$. Also, the change in age performance is more severe in the high interference case ($K = 5$ or low $K$). Choosing $\beta = 0$ appears to be the safest bet for the age-based policy $\pi_A$.

\section{Conclusion}
\label{sec:conclusion}
We considered the problem of age minimization for a wireless network under general interference constraints, and time varying channel. We proposed two policies: virtual queue based policy and age-based policy. We proved that the virtual queue based policy is peak age optimal, barring an additive factor, while the age-based policy is at most factor $4$ away from the optimal peak and average age. Using numerical simulations, we demonstrated that both the proposed policies are, in fact, very close to optimal.


\bibliographystyle{ieeetr}


\appendix

\subsection{Proof of Lemma~\ref{lem:time_conservation}}
\label{pf:lem:time_conservation}
Age evolution for link $e$, is given by (see~\eqref{eq:age_evol2}):
\begin{equation}
A_{e}(t+1) = 1 + A_{e}(t) - U_{e}(t)S_{e}(t)A_{e}(t),
\end{equation}
for all $t$. Summing this over $t$ time slots we obtain
\begin{align}
A_{e}(t) - A_{e}(0) &= \sum_{\tau = 0}^{t-1} \left( A_{e}(\tau + 1) - A_{e}(\tau)\right), \nonumber \\
&= \sum_{\tau = 0}^{t-1} \left( 1 - U_{e}(\tau)S_{e}(\tau)A_{e}(\tau)\right), \nonumber \\
&= t -  \sum_{\tau = 0}^{t-1}U_{e}(\tau)S_{e}(\tau)A_{e}(\tau).
\end{align}
Taking expected value on both sides yields
\begin{equation}
\label{eq:t1}
\frac{1}{t}\EX{A_{e}(t)} - \frac{1}{t}\EX{A_{e}(0)} = 1 - \frac{1}{t} \EX{\sum_{\tau = 0}^{t-1}U_{e}(\tau)S_{e}(\tau)A_{e}(\tau)}.
\end{equation}
Note that, for $\pi \in \Pi$, we have a bounded $\EX{A_{e}(t)}$ for all $t$, and $\limsup_{t \rightarrow \infty} \frac{1}{t}\EX{A_{e}(t)} = 0$. This follows from ergodicity of the process $\{ S_{e}(t)U_{e}(t) \}_{t \geq 0}$. Note that $\{ S_{e}(t)U_{e}(t) \}_{t \geq 0}$ is ergodic because $\{ U_{e}(t)\}_{t\geq 0}$ is ergodic (due to~\eqref{eq:Pi}) and $S_{e}(t)$ is i.i.d. across time $t$ and independent of $U_{e}(t)$. Taking $t \rightarrow \infty$ in~\eqref{eq:t1}, and using $\limsup_{t \rightarrow \infty} \frac{1}{t}\EX{A_{e}(t)} = 0$, yields the result.

\subsection{Derivation of the Peak Age Minimization Problem}
\label{pf:der_peak_min_prob}
Using Lemma~\ref{lem:time_conservation}, we first show that for $\pi \in \Pi$ is given by
\begin{equation}
\overline{A}^{\text{p}}_{e} = \frac{1}{\liminf_{t \rightarrow \infty} \EX{\frac{1}{t}\sum_{\tau = 0}^{t}\sum_{e \in E}U_{e}(\tau)S_{e}(\tau)}},
\end{equation}
for every $e \in E$. By definition, the peak age of link $e$ is given by
\begin{align}
\overline{A}^{\text{p}}_{e} &= \limsup_{t \rightarrow \infty} \frac{ \EX{\sum_{\tau = 0}^{t-1} U_{e}(\tau)S_{e}(\tau)A_{e}(\tau)} }{ \EX{\sum_{\tau = 0}^{t-1}U_{e}(\tau)S_{e}(\tau)} }, \nonumber \\
&= \limsup_{t \rightarrow \infty} \frac{  \EX{\frac{1}{t}\sum_{\tau = 0}^{t-1} U_{e}(\tau)S_{e}(\tau)A_{e}(\tau)} }{ \EX{\frac{1}{t}\sum_{\tau = 0}^{t-1}U_{e}(\tau)S_{e}(\tau)} }, \nonumber \\
&= \frac{ \limsup_{t\rightarrow \infty} \EX{\frac{1}{t}\sum_{\tau = 0}^{t-1} U_{e}(\tau)S_{e}(\tau)A_{e}(\tau)} }{ \liminf_{t \rightarrow \infty} \EX{\frac{1}{t}\sum_{\tau = 0}^{t-1}U_{e}(\tau)S_{e}(\tau)} }, \nonumber \\
&= \frac{1}{ \liminf_{t \rightarrow \infty} \EX{\frac{1}{t}\sum_{\tau = 0}^{t-1}U_{e}(\tau)S_{e}(\tau)} }, \label{eq:tani}
\end{align}
where the last equality follows from Lemma~\ref{lem:time_conservation}. Since $\overline{A}^{\text{p}}(\pi) = \sum_{e \in E}w_e \overline{A}^{\text{p}}_{e}(\pi)$, the peak age minimization problem $\min_{\pi \in \Pi} \overline{A}^{\text{p}}(\pi)$ can now be written as
\begin{align}
\begin{aligned}
& \underset{\pi \in \Pi}{\text{Minimize}} & & \sum_{e \in E}\frac{w_e}{\liminf_{t \rightarrow \infty} \frac{1}{t}\sum_{\tau = 0}^{t-1}U_{e}(\tau)S_{e}(\tau)}.
\end{aligned}
\end{align}
Using auxiliary variables $\alpha_e$, this can be written as~\eqref{eq:peak_min_problem}.

\subsection{Proof of Theorem~\ref{thm:piQ}}
\label{pf:thm:piQ}
\emph{Proof of Part~A}: Let $L(t) = \frac{1}{2}\sum_{e \in E}w_e Q^{2}_{e}(t)$ and $\Delta(t) = L(t+1) - L(t)$. Note that
\begin{align}
Q_{e}^{2}(t+1) &= \left[ \max\{ Q_{e}(t) + \alpha_{e}(t) - U_{e}(t)S_{e}(t), 1\} \right]^2, \nonumber \\
&\leq 1 + \left( Q_{e}(t) + \alpha_{e}(t) - U_{e}(t)S_{e}(t) \right)^2, \nonumber \\
&= 1 + \left( \alpha_{e}(t) - U_{e}(t)S_{e}(t) \right)^2 + Q^{2}_{e}(t) \nonumber \\
&~~~~~~~~~~~~~~~~~~+ 2Q_{e}(t)\left( \alpha_{e}(t) - U_{e}(t)S_{e}(t) \right), \nonumber \\
&\leq 1 + V + Q^{2}_{e}(t) + 2Q_{e}(t)\left( \alpha_{e}(t) - U_{e}(t)S_{e}(t) \right), \label{eq:r3}
\end{align}
where the last inequality follows from the fact that $\alpha_{e}(t) = \sqrt{\frac{V}{Q_{e}(t)}} \leq \sqrt{V}$ because $Q_{e}(t) \geq 1$ for all $t$.
Using~\eqref{eq:r3} we obtain
\begin{equation}
\label{eq:r4}
\Delta(t) \leq \frac{1+V}{2}\sum_{e \in E}w_e + \sum_{e \in E}w_e Q_{e}(t)\left( \alpha_{e}(t) - U_{e}(t)S_{e}(t) \right),
\end{equation}
for all $t$. We, therefore, have
\begin{multline}
Vg(\bm{\alpha}(t)) + \Delta(t) \leq V \sum_{e \in E}\frac{w_e}{\alpha_e(t)} + \frac{1+V}{2}\sum_{e \in E}w_e \\
+ \sum_{e \in E}w_eQ_{e}(t)\left[ \alpha_{e}(t) - U_{e}(t)S_{e}(t)\right]. \nonumber
\end{multline}
Substituting $\alpha_{e}(t) = \sqrt{V/Q_{e}(t)}$, we get
\begin{multline}
Vg(\bm{\alpha}(t)) + \Delta(t) \leq \sum_{e \in E}2w_e\sqrt{VQ_{e}(t)} \\ + \frac{1+V}{2}\sum_{e \in E}w_e
- \sum_{e \in E}w_e U_{e}(t)S_{e}(t)Q_{e}(t). \label{eq:kk4}
\end{multline}
Taking conditional expectation, with respect to $\mathbf{U}(t)$ and $\mathbf{Q}(t)$, we obtain
\begin{multline}
\EX{Vg(\bm{\alpha}(t)) + \Delta(t) | \mathbf{U}(t), \mathbf{Q}(t)} \leq \sum_{e \in E}2w_e\sqrt{VQ_{e}(t)} \\ + \frac{1+V}{2}\sum_{e \in E}w_e
- \sum_{e \in E}w_e \gamma_e U_{e}(t)Q_{e}(t), \label{eq:kk4x}
\end{multline}
since $S_{e}(t)$ is i.i.d. across time.
The policy $\pi_Q$ minimizes the right hand side of~\eqref{eq:kk4x}, as it activates set $m_{t}$ at $t$ which maximizes $\sum_{e \in m}w_e \gamma_e Q_{e}(t)$. Therefore, we can upper bound the right-hand side of~\eqref{eq:kk4x} by the peak age optimal stationary policy $\pi_C$:
\begin{multline}
\EX{ Vg(\bm{\alpha}(t)) + \Delta(t) | \mathbf{U}(t), \mathbf{Q}(t)} \leq \sum_{e \in E}2w_e\sqrt{VQ_{e}(t)} \\ + \frac{1+V}{2}\sum_{e \in E}w_e - \sum_{e \in E}w_e \gamma_e U^{\pi_C}_{e}(t)Q_{e}(t). \label{eq:kk4xx}
\end{multline}
Note that the link activation frequency of the policy $\pi_C$ is $f^{\ast}_e = \EX{U^{\pi_C}_{e}(t)}$, where $f^{\ast}_{e}$ is the solution to the problem~\eqref{eq:peak_age_opt}. Taking expectation with decision variables $\mathbf{U}(t)$ and $\mathbf{U}^{\pi_C}(t)$ we get
\begin{multline}
\EX{Vg(\bm{\alpha}(t)) + \Delta(t)|\mathbf{Q}(t)} \leq \sum_{e \in E}2w_e\sqrt{VQ_{e}(t)} \\ + \frac{1+V}{2}\sum_{e \in E}w_e
- \sum_{e \in E}w_e \gamma_e  f^{\ast}_{e} Q_{e}(t). \label{eq:kk5}
\end{multline}
This can be written as
\begin{multline}
\EX{Vg(\bm{\alpha}(t)) + \Delta(t)|\mathbf{Q}(t)} \leq V \overline{A}^{\text{p}\ast} + \frac{1+V}{2}\sum_{e \in E}w_e \\
- \sum_{e \in E}w_e \gamma_e f^{\ast}_{e}\left[ \sqrt{Q_{e}(t)} - \frac{\sqrt{V}}{\gamma_e f^{\ast}_{e}} \right]^2, \label{eq:kkk20}
\end{multline}
where $\overline{A}^{\text{p}\ast} = \sum_{e \in E}\frac{w_e}{\gamma_e f^{\ast}_{e}}$ is the optimal value given in~\eqref{eq:peak_age_opt}.

Now, ignoring the last term in~\eqref{eq:kkk20}, taking expected value, and summing both sides of~\eqref{eq:kkk20} over the first $t$ time slots we obtain
\begin{multline}
\EX{V \sum_{\tau=0}^{t-1} g(\bm{\alpha}(t))} + \EX{L(t) - L(0)} \\
\leq t\left[ V \overline{A}^{\text{p}\ast} + \frac{1+V}{2}\sum_{e \in E}w_e\right]. \nonumber
\end{multline}
Since $L(t) \geq 0$, we have
\begin{align}
\EX{V \sum_{\tau=0}^{t-1} g(\bm{\alpha}(t))} &\leq \EX{V \sum_{\tau=0}^{t-1} g(\bm{\alpha}(t))} + \EX{L(t)}, \nonumber \\
&\leq t\left[V \overline{A}^{\text{p}\ast} + \frac{1+V}{2}\sum_{e \in E}w_e\right] + \EX{L(0)}. \nonumber
\end{align}
Diving by $t$ and $V$, and taking the limit $\limsup_{t \rightarrow \infty}$, we get
\begin{equation}
\limsup_{t \rightarrow \infty} \frac{1}{t}\EX{\sum_{\tau = 0}^{t-1}g(\bm{\alpha}(t))} \leq \overline{A}^{\text{p}\ast} + \frac{1}{2}\sum_{e \in E}w_e +  \frac{1}{2V}\sum_{e \in E}w_e. \label{eq:kk6}
\end{equation}
Since $g$ is convex, we have $g(\overline{\bm{\alpha}}(t)) \leq \frac{1}{t}\sum_{\tau = 0}^{t-1}g(\bm{\alpha}(t))$ from Jensen's inequality~\cite{boyd}. Substituting this in~\eqref{eq:kk6} yields the result:
\begin{equation}
\limsup_{t \rightarrow \infty} \EX{g(\overline{\bm{\alpha}}(t))} \leq \overline{A}^{\text{p}\ast} + \frac{1}{2}\sum_{e \in E}w_e +  \frac{1}{2V}\sum_{e \in E}w_e.
\end{equation}

\emph{Proof of Part~B}: Since $V g(\bm{\alpha}(t)) \geq 0$, from~\eqref{eq:kkk20} we obtain
\begin{equation}
\EX{\Delta(t)} \leq V\left[ \overline{A}^{\text{p}\ast} + \frac{1}{2}\sum_{e \in E}w_e \right] + \frac{1}{2}\sum_{e \in E}w_e.
\end{equation}
Summing this over $t$ time slots we get
\begin{equation}
\frac{1}{t}\EX{L(t)} \leq \frac{1}{t}\EX{L(0)} + V\left[ \overline{A}^{\text{p}\ast} + \frac{1}{2}\sum_{e \in E}w_e \right] + \frac{1}{2}\sum_{e \in E}w_e.
\end{equation}
This implies,
\begin{equation}
\label{eq:nd1}
\limsup_{t \rightarrow \infty} \frac{1}{t}\EX{L(t)} \leq B,
\end{equation}
where $B = V\left[ \overline{A}^{\text{p}\ast} + \frac{1}{2}\sum_{e \in E}w_e \right] + \frac{1}{2}\sum_{e \in E}w_e$. Now, since $L(t) = \frac{1}{2}\sum_{e \in E}w_e Q^{2}_{e}(t)$,~\eqref{eq:nd1} implies
\begin{equation}
\limsup_{t \rightarrow \infty} \frac{1}{t}\EX{Q^{2}_{e}(t)} \leq B,
\end{equation}
and as a consequence  $\limsup_{t \rightarrow \infty} \frac{1}{\sqrt{t}}\EX{Q_{e}(t)} \leq B$, for all $e \in E$, since $\EX{Q_{e}(t)}^2 \leq \EX{Q^2_{e}(t)}$. This implies
\begin{equation}
\limsup_{t \rightarrow \infty} \frac{1}{t}\EX{Q_{e}(t)} = 0,
\end{equation}
for all $e \in E$.

\emph{Proof of Part~C}: The queue evolution equation implies
\begin{equation}
Q_{e}(\tau+1) \geq Q_{e}(\tau) + \alpha_{e}(\tau) - U_{e}(\tau)S_{e}(\tau),
\end{equation}
for any $\tau \geq 0$. Summing this over $t$ times slots yields
\begin{equation}
\label{eq:w1}
\overline{\alpha}_{e}(t) + \frac{1}{t}Q_{e}(0) \leq \frac{1}{t}\sum_{\tau=0}^{t-1}U_{e}(\tau)S_{e}(\tau) + \frac{1}{t}Q_{e}(t),
\end{equation}
for all $t \geq 0$. Since $Q_{e}(t)$ is mean rate stable, taking expected value of~\eqref{eq:w1} and liminf as $t \rightarrow \infty$ we obtain
\begin{equation}
\liminf_{t \rightarrow \infty} \EX{\overline{\alpha}_{e}(t)} \leq \liminf_{t \rightarrow \infty} \frac{1}{t}\EX{\sum_{\tau=0}^{t-1}U_{e}(\tau)S_{e}(\tau)}.
\label{eq:w2}
\end{equation}
Since, $S_{e}(t)$ is independent of $U_{e}(t)$,~\eqref{eq:w2} implies the result:
\begin{equation}
\frac{1}{\gamma_e}\liminf_{t \rightarrow \infty} \EX{\overline{\alpha}_{e}(t)} \leq \liminf_{t \rightarrow \infty} \frac{1}{t}\EX{\sum_{\tau=0}^{t-1}U_{e}(\tau)}.
\label{eq:w2}
\end{equation}

Furthermore, since $g$ is a continuous, decreasing function in each $\alpha_e$ we have
\begin{align}
\overline{A}^{\text{p}}(\pi_Q) &= \sum_{e \in E} \frac{w_e}{\liminf_{t \rightarrow \infty} \EX{\frac{1}{t}\sum_{\tau = 0}^{t-1}U_{e}(t)S_{e}(t)}}, \nonumber \\
&\leq \sum_{e \in E} \frac{w_e}{\liminf_{t \rightarrow \infty} \EX{\overline{\alpha}_{e}(t)}}, \nonumber \\
&= \limsup_{t \rightarrow \infty} \sum_{e \in E} \frac{w_e}{\EX{\overline{\alpha}_{e}(t)}}, \nonumber \\
&\leq  \limsup_{t \rightarrow \infty} \EX{\sum_{e \in E} \frac{w_e}{\overline{\alpha}_{e}(t)} } = \limsup_{t \rightarrow \infty} \EX{g\left( \overline{\bm{\alpha}}(t)\right) },\label{eq:kk2}
\end{align}
where the first equality follows from Lemma~\ref{lem:time_conservation} and~\eqref{eq:peak_def}, the second inequality follows from~\eqref{eq:w2}, while the last inequality follows from Jensen's inequality~\cite{Durrett} and definition of $g(\bm{\alpha})$.

\subsection{Proof of Lemma~\ref{lem:ave_age_conservation}}
\label{pf:lem:ave_age_conservation}
The age of link $e$ evolves as (see~\eqref{eq:age_evol2}):
\begin{equation}
A_{e}(t+1) = 1 + A_{e}(t) - U_{e}(t)S_{e}(t)A_{e}(t),
\end{equation}
for all $t$. Squaring this we obtain
\begin{multline}
\label{eq:s1}
A^{2}_{e}(t+1) = 1 + A^{2}_{e}(t) + U^{2}_{e}(t)S^{2}_{e}(t)A^{2}_{e}(t) + 2A_{e}(t) \\ - 2U_{e}(t)S_{e}(t)A^{2}_{e}(t) - 2U_{e}(t)S_{e}(t)A_{e}(t).
\end{multline}
Since $U_{e}(t)S_{e}(t) \in \{0, 1\}$, we have $U^{2}_{e}(t)S^{2}_{e}(t) = U_{e}(t)S_{e}(t)$. Substituting this in~\eqref{eq:s1} we get
\begin{multline}\label{eq:s2}
A^{2}_{e}(t+1) - A^{2}_{e}(t) = 1 + 2A_{e}(t) - U_{e}(t)S_{e}(t)A^{2}_{e}(t) \\ - 2U_{e}(t)S_{e}(t)A_{e}(t),
\end{multline}
for all $t$. Telescoping this over $t$ time slots we get
\begin{align}
A^{2}_{e}(t) - A^{2}_{e}(0) &= \sum_{\tau = 0}^{t-1} \left( A^{2}_{e}(\tau + 1) - A^{2}_{e}(\tau) \right), \nonumber \\
&= t + 2 \sum_{\tau = 0}^{t-1} A_{e}(\tau) -  \sum_{\tau = 0}^{t-1}U_{e}(\tau)S_{e}(\tau)A^{2}_{e}(\tau) \nonumber \\
&~~~~~~~~~- 2 \sum_{\tau = 0}^{t-1}U_{e}(\tau)S_{e}(\tau)A_{e}(\tau). \label{eq:s3}
\end{align}

For a policy $\pi \in \Pi$, we must have $\limsup_{t \rightarrow \infty} \frac{1}{t}\EX{A^{2}_{e}(t)} = 0$. This follows from ergodicity of the process $\{ S_{e}(t)U_{e}(t) \}_{t \geq 0}$. Note that $\{ S_{e}(t)U_{e}(t) \}_{t \geq 0}$ is ergodic because $\{ U_{e}(t)\}_{t\geq 0}$ is ergodic (due to~\eqref{eq:Pi}) and $S_{e}(t)$ is i.i.d. across time $t$ and independent of $U_{e}(t)$. Taking expectation in~\eqref{eq:s3}, using $\frac{1}{t}\EX{A^{2}_{e}(t)} \rightarrow 0$, we get
\begin{align}
2\overline{A}^{\text{ave}}_{e} &= -1 + \limsup_{t \rightarrow \infty} \frac{1}{t}\EX{\sum_{\tau = 0}^{t-1}U_{e}(\tau)S_{e}(\tau)A^{2}_{e}(\tau)} \nonumber \\
&~~~~~+ 2\limsup_{t \rightarrow \infty} \frac{1}{t}\EX{\sum_{\tau=0}^{t-1}U_{e}(\tau)S_{e}(\tau)A_{e}(\tau)}, \nonumber \\
&= 1 + \limsup_{t \rightarrow \infty} \frac{1}{t}\EX{\sum_{\tau = 0}^{t-1}U_{e}(\tau)S_{e}(\tau)A^{2}_{e}(\tau)}, \label{eq:nnn333}
\end{align}
where the last equality follows from Lemma~\ref{lem:time_conservation}. This proves the Lemma for $\beta = 0$. From Lemma~\ref{lem:time_conservation}, we have that
\begin{equation}
0 = - 1 + \limsup_{t \rightarrow \infty} \frac{1}{t}\EX{\sum_{\tau = 0}^{t-1}U_{e}(\tau)S_{e}(\tau)A_{e}(\tau)}. \label{eq:nex}
\end{equation}
Adding $\beta$ times~\eqref{eq:nex} to~\eqref{eq:nnn333} we obtain the result, for any $\beta \in \mathbb{R}$. Note that in~\eqref{eq:nnn333} and~\eqref{eq:nex}, $S_{e}(\tau)$ can be replaced by $\gamma_e$ because $S_{e}(\tau)$ is independent of $U_{e}(\tau)$ and $A_{e}(\tau)$.

\subsection{Proof of Theorem~\ref{thm:piA}}
\label{pf:thm:piA}
Define $L(t) = \frac{1}{2}\sum_{e \in E} w_e A_{e}^{2}(t)$, $\Delta(t) = L(t+1) - L(t)$, and
\begin{multline}
\label{eq:q1}
f(t) =  \left( 1 - \beta\frac{(1-V)}{2}\right)\sum_{e \in E} w_e U_{e}(t)S_{e}(t)A_{e}(t) \\
+ \frac{V}{2}\sum_{e \in E} w_e U_{e}(t) S_{e}(t) A_{e}^{2}(t),
\end{multline}
for $0 < V < 1$, $\beta \in \mathbb{R}$, and all $t \geq 0$.
Using age evolution equation $A_{e}(t+1) = 1 + A_{e}(t) - U_{e}(t)S_{e}(t)A_{e}(t)$,
we obtain
\begin{multline}
\label{eq:q2}
\Delta(t) = \frac{1}{2}\sum_{e \in E}w_e + \sum_{e \in E}w_e A_{e}(t) \\
-\sum_{e \in E}w_e U_{e}(t)S_{e}(t)A_{e}(t) - \frac{1}{2}\sum_{e \in E}w_e U_{e}(t)S_{e}(t)A^{2}_{e}(t).
\end{multline}
Summing~\eqref{eq:q1} and~\eqref{eq:q2} we get
\begin{multline}
f(t) + \Delta(t) = \frac{1}{2}\sum_{e \in E}w_e + \sum_{e \in E} w_e A_{e}(t) \\ - \frac{(1-V)}{2}\sum_{e \in E}w_e U_{e}(t) S_{e}(t) \left[ A_{e}^{2}(t) + \beta A_{e}(t) \right],
\end{multline}
and taking conditional expectation, we obtain
\begin{multline}
\EX{f(t) + \Delta(t) \big| \mathbf{U}(t), \mathbf{A}(t) }= \frac{1}{2}\sum_{e \in E}w_e + \sum_{e \in E} w_e A_{e}(t) \\ - \frac{(1-V)}{2}\sum_{e \in E}w_e \gamma_e U_{e}(t) \left[ A_{e}^{2}(t) + \beta A_{e}(t) \right].
\label{eq:nnn}
\end{multline}
The policy $\pi_A$ chooses $\mathbf{U}(t)$ that maximizes
\begin{equation}
\sum_{e \in E}w_e \gamma_e U_{e}(t) \left[ A_{e}^{2}(t) + \beta A_{e}(t) \right],
\end{equation}
and thus, it minimizes the right-hand side in~\eqref{eq:nnn}. Therefore, for any other policy $\pi$, we must have
\begin{multline}
\EX{f(t) + \Delta(t) \big| \mathbf{U}(t), \mathbf{A}(t) } \leq \frac{1}{2}\sum_{e \in E}w_e + \sum_{e \in E} w_e A_{e}(t) \\ - \frac{(1-V)}{2}\sum_{e \in E}w_e \gamma_e U^{\pi}_{e}(t) \left[ A_{e}^{2}(t) + \beta A_{e}(t)\right],
\end{multline}
where $\mathbf{U}^{\pi}(t)$ denotes the action of policy $\pi$ at time $t$. Substituting $\pi = \pi_C$, which is the stationary peak age optimal policy that solves~\eqref{eq:peak_age_opt}, gives the bound
\begin{multline}
\EX{f(t) + \Delta(t) \big| \mathbf{A}(t)} \leq \frac{1}{2}\sum_{e \in E}w_e + \sum_{e \in E} w_e A_{e}(t) \\
- \frac{(1-V)}{2}\sum_{e \in E}w_e \gamma_e f^{\ast}_e\left[A_{e}^{2}(t) + \beta A_{e}(t)\right], \label{eq:last_one}
\end{multline}
This can be re-written as
\begin{multline}
\EX{f(t) + \Delta(t) \big| \mathbf{A}(t)} \leq \frac{1}{2}\sum_{e \in E}w_e\\
+ \frac{1-V}{2}\sum_{e \in E} w_e \gamma_e f^{\ast}_{e} \left[ \frac{\beta^2}{4} + \frac{(1-V)^{-2}}{\gamma^{2}_e f^{\ast 2}_{e}} - \frac{1}{1-V}\frac{\beta}{\gamma_e f^{\ast}_e }\right]\\
- \frac{(1-V)}{2}\!\sum_{e \in E}w_e \gamma_e f^{\ast}_e \! \left[ A_{e}(t) + \frac{\beta}{2} - \frac{(1-V)^{-1}}{\gamma_e f^{\ast}_{e} }\right]^2. \label{eq:star0}
\end{multline}
Ignoring the last term, since it is negative, and using the fact that $\gamma_e f^{\ast}_{e} \leq 1$ we have
\begin{multline}
\EX{f(t) + \Delta(t)} \leq \frac{(1-V)^{-1}}{2}\sum_{e \in E}\frac{w_e}{\gamma_e f^{\ast}_e} + \theta \sum_{e \in E}w_e,
\label{eq:star}
\end{multline}
where $\theta = \frac{1-\beta}{2} + (1-V)\frac{\beta^2}{4}$.
Summing this over $t$ time slots we obtain
\begin{multline}
\EX{\sum_{\tau = 0}^{t-1}f(\tau)} + \EX{L(t) - L(0)} \\
\leq t \left[ \frac{(1-V)^{-1}}{2}\sum_{e \in E}\frac{w_e}{\gamma_e f^{\ast}_e} + \theta \sum_{e \in E}w_e \right].
\end{multline}
Since $L(t) \geq 0$ for all $t$, we have
\begin{align}
&\EX{\sum_{\tau = 0}^{t-1}f(\tau)} \leq  \EX{\sum_{\tau = 0}^{t-1}f(\tau)} + \EX{L(t)}, \nonumber  \\
&~~~~~~\leq t \left[ \frac{(1-V)^{-1}}{2}\sum_{e \in E}\frac{w_e}{\gamma_e f^{\ast}_e} + \theta \sum_{e \in E}w_e \right] + \EX{L(0)}. \nonumber
\end{align}
Dividing this by $t$ and taking the limit we obtain
\begin{equation}
\limsup_{t \rightarrow  \infty} \frac{1}{t} \EX{\sum_{\tau = 0}^{t-1}f(\tau)} \leq \frac{(1-V)^{-1}}{2}\sum_{e \in E}\frac{w_e}{\gamma_e f^{\ast}_e} + \theta \sum_{e \in E}w_e. \label{eq:star2}
\end{equation}
Note that $\overline{A}^{\text{p}\ast} = \sum_{e \in E}\frac{w_e}{\gamma_e f^{\ast}_e}$, from~\eqref{eq:peak_age_opt} in Theorem~\ref{thm:stationary}. Further, we also know from Theorem~\ref{thm:stationary} that $\overline{A}^{\text{p}\ast} \leq 2\overline{A}^{\text{ave}\ast} - \sum_{e \in E} w_e$. Substituting this in~\eqref{eq:star2} we get
\begin{multline}
\limsup_{t \rightarrow \infty} \EX{\frac{1}{t}\sum_{\tau=0}^{t-1}f(\tau)} \leq \frac{1}{(1-V)}\overline{A}^{\text{ave}\ast} \\
+ \left( \theta - \frac{1}{2(1-V)}\right)\sum_{e \in E} w_e. \label{eq:star3}
\end{multline}

Assuming that $\EX{A_{e}^{2}(t)}$ is uniformly bounded for all $t$, we can make use of Lemma~\ref{lem:time_conservation} and~\ref{lem:ave_age_conservation} to compute $\limsup_{t \rightarrow \infty} \EX{\frac{1}{t}\sum_{\tau=0}^{t-1}f(\tau)}$. This gives us
\begin{multline}
\limsup_{t \rightarrow \infty} \EX{\frac{1}{t}\sum_{\tau=0}^{t-1}f(\tau)} = \sum_{e \in E} w_e + V\overline{A}^{\text{ave}}\left( \pi_A\right) \\
- \frac{\beta(1-V) + V}{2}\sum_{e \in E} w_e. \label{eq:nnt}
\end{multline}
Substituting this in~\eqref{eq:star3} we get
\begin{equation}
\overline{A}^{\text{ave}}\left( \pi_A\right) \leq \frac{1}{V(1-V)}\overline{A}^{\text{ave}\ast} - \kappa \sum_{e \in E} w_e,
\end{equation}
where $\kappa$ is given by
\begin{equation}
\kappa = \frac{1}{V} + \frac{1}{2V(1-V)} - \frac{\beta(1-V) + V}{2V} - \frac{\theta}{V}.
\end{equation}
Substituting $V = 1/2$ gives the result in~\eqref{eq:thm:piA_1}.

In order to obtain~\eqref{eq:thm:piA_2}, notice that~\eqref{eq:star2} can be written as
\begin{equation}
\label{eq:nnt10}
\limsup_{t \rightarrow  \infty} \frac{1}{t} \EX{\sum_{\tau = 0}^{t-1}f(\tau)} \leq \frac{(1-V)^{-1}}{2}\overline{A}^{\text{p}\ast} + \theta \sum_{e \in E}w_e,
\end{equation}
since $\overline{A}^{\text{p}\ast} = \sum_{e \in E}\frac{w_e}{\gamma_e f^{\ast}_{e}}$, see Theorem~\ref{thm:stationary}. We state and use the following result from~\cite{talak17_StGenIC_Mobihoc}:
\begin{framed}
\begin{lemma}
\label{lem:peak_ave_bound}
For any policy $\pi \in \Pi$, we have
\begin{equation}
\overline{A}^{\text{p}}(\pi) \leq 2\overline{A}^{\text{ave}}(\pi) - \sum_{e \in E}w_e.
\end{equation}
\end{lemma}
\end{framed}
Now, using~\eqref{eq:nnt} and $\overline{A}^{\text{p}}(\pi_A) \leq 2\overline{A}^{\text{ave}}(\pi_A) - \sum_{e \in E}w_e$, from Lemma~\ref{lem:peak_ave_bound}, we get
\begin{multline}
\label{eq:nnt11}
\limsup_{t \rightarrow  \infty} \frac{1}{t} \EX{\sum_{\tau = 0}^{t-1}f(\tau)} \geq \sum_{e \in E} w_e + \frac{V}{2}\overline{A}^{\text{p}}\left( \pi_A\right) \\
- \frac{\beta(1-V)}{2}\sum_{e \in E} w_e.
\end{multline}
Combining~\eqref{eq:nnt10} and~\eqref{eq:nnt11} in order to obtain a bound on $\overline{A}^{\text{p}}\left( \pi_A\right)$ as a function of $\overline{A}^{\text{p}\ast}$, and setting $V = 1/2$, we get the result in~\eqref{eq:thm:piA_2}.

It suffices to argue that the mean $\EX{A^{2}_{e}(t)}$ is uniformly bounded for all $t$, for policy $\pi_A$.\footnote{Unlike in the proof of Lemma~\ref{lem:ave_age_conservation}, we do not know this in advance as we have not yet shown that $\pi_A \in \Pi$. Positive recurrence of the process $\{ \mathbf{A}(t)\}_{t \geq 0}$, proved here, establishes this fact.} Define a Lyapunov function $\tilde{L}(t) = \frac{1}{2}\sum_{e \in E}w_e \left( A_{e}(t) + \beta/2 - 1\right)^2$, and the corresponding drift $\tilde{\Delta}(t) = \tilde{L}(t+1)  - \tilde{L}(t)$. Then using the same arguments as in~\eqref{eq:star0} we can obtain
\begin{equation}
\EX{\tilde{\Delta}(t)|\mathbf{A}(t)} \leq B_1 - \sum_{e \in E}B_{2,e}\left(A_{e}(t) + c_e \right)^{2},
\end{equation}
for constants $B_1$, $B_{2,e}$, and $c_e$. Foster-Lyapunov theorem~\cite[Chap.~6]{Hajek_rp} then implies that the process $\{ \mathbf{A}^{2}(t) \}_{t}$ is positive recurrent, and that $\EX{A^{2}_{e}(t)}$ is uniformly bounded.

\end{document}